\documentclass[11pt,a4paper]{article}
\usepackage{jcappub}

\usepackage{bm}
\usepackage{amsfonts}
\usepackage{subfigure}

\renewcommand\({\left(}
\renewcommand\){\right)}
\renewcommand\[{\left[}
\renewcommand\]{\right]}

\newcommand{\be}{\begin{equation}}
\newcommand{\ee}{\end{equation}}
\newcommand{\bea}{\begin{eqnarray}}
\newcommand{\eea}{\end{eqnarray}}

\newcommand{\fabs}[1]{\left| #1 \right|}
\newcommand{\para}[1]{\left( #1 \right)}
\newcommand{\di}{\text{d}}
\renewcommand{\vec}[1]{{\bm #1}}
\begin{document}


\subheader{\hfill MPP-2012-79}

\title{Thermalisation of light sterile neutrinos in the early universe}

\author[a]{Steen Hannestad,}
\author[b]{Irene Tamborra}
\author[a]{and Thomas Tram}

\affiliation[a]{Department of Physics and Astronomy,
 University of Aarhus, 8000 Aarhus C, Denmark}

\affiliation[b]{Max-Planck-Institut f\"ur Physik
(Werner-Heisenberg-Institut)\\
F\"ohringer Ring 6, 80805 M\"unchen, Germany}

\emailAdd{sth@phys.au.dk}
\emailAdd{tamborra@mpp.mpg.de}
\emailAdd{tram@phys.au.dk}

\abstract{Recent  cosmological data favour additional relativistic degrees of freedom beyond  the three active
 neutrinos and photons, often referred to as ``dark'' radiation. Light sterile neutrinos is one of the
 prime candidates for such additional radiation. However, constraints on sterile neutrinos based on the current cosmological data
  have been derived using simplified assumptions about thermalisation of $\nu_s$ at the Big Bang Nucleosynthesis (BBN) epoch.
  These assumptions are not necessarily justified and here we solve the full quantum kinetic equations in the ($1$ active + $1$ sterile) scenario and
  derive the number of thermalised species just before BBN begins ($T \simeq 1$~MeV) for null ($L=0$) and
  large ($L=10^{-2}$) initial lepton asymmetry and for a range of possible mass-mixing parameters. We find that the
  full thermalisation assumption during the BBN epoch is justified for initial small lepton asymmetry only. Partial or null thermalisation
  occurs when the initial lepton asymmetry is large.}
\maketitle

\section{Introduction}                        \label{sec:introduction}
Sterile neutrinos are hypothetical $SU(2) \times U(1)$ singlets. They are supposed to
mix with one or more of the active states without interacting with any other particle.
Low-mass sterile neutrinos have been invoked to explain the excess $\bar{\nu}_e$ events in the LSND
experiment~\cite{Aguilar:2001ty,Strumia:2002fw,GonzalezGarcia:2007ib} as well
as the MiniBooNE excess events in both neutrino and antineutrino
channels. Interpreted in terms of flavour oscillations, the MiniBooNE
data require CP violation and thus no less than two sterile
families~\cite{AguilarArevalo:2008rc,AguilarArevalo:2009xn,Karagiorgi:2009nb} or
 additional ingredients such as non-standard interactions~\cite{Akhmedov:2010vy}.
  Recently  a new analysis of reactor $\bar\nu_e$ spectra
and their distance and energy variation~\cite{Mention:2011rk,Huber:2011wv,Kopp:2011qd} suggested
indication for the possible existence of eV-mass sterile neutrinos. However the  IceCube
collaboration excluded part of the parameter space~\cite{Razzaque:2011ab}.

The most recent analysis of  cosmological data suggest  a trend towards the
existence of ``dark radiation,'' radiation in excess with respect to the three neutrino families
and photons~\cite{Hamann:2007pi,Hamann:2010pw,GonzalezGarcia:2010un}. The
cosmic radiation content is usually expressed in terms of the effective number of thermally excited
neutrino species, $N_{\rm eff}$. Its standard value, $N_{\rm eff}=3.046$,
slightly exceeds 3 because of $e^+e^-$ annihilation providing residual
neutrino heating~\cite{Mangano:2005cc}. The Wilkinson
Microwave Anisotropy Probe (WMAP) collaboration found
$N_{\rm eff} = 4.34^{+0.86}_{-0.88}$ based on their 7-year
data release and additional LSS data~\cite{Komatsu:2010fb} at $1 \sigma$.
 Including the Sloan Digital Sky Survey (SDSS) data release 7
(DR7) halo power spectrum, \cite{Hamann:2010pw} found $N_{\rm eff} = 4.78^{+1.86}_{-1.79}$
at $2 \sigma$.
Measurements of the CMB anisotropy on smaller scales by the ACT \cite{Dunkley:2010ge} and SPT \cite{Keisler:2011aw}
collaborations also find
tentative evidence for a value of $N_{\rm eff}$ higher than predicted by the standard model (see also
\cite{Hamann:2011hu,Joudaki:2012fx,Giusarma:2011zq,Nollett:2011aa,GonzalezMorales:2011ty}
for recent discussions of $N_{\rm eff}$).

Also, cosmological constraints coming from big bang nucleosynthesis (BBN) suggest that the relatively high
 $^4$He abundance can be interpreted in terms of  additional
 radiation during the BBN epoch~\cite{Izotov:2010ca,Aver:2010wq}.
Low-mass sterile neutrinos have been considered among possible candidates for the
extra-radiation content~\cite{Hamann:2011ge,Hamann:2010bk,Giusarma:2011ex}.
The cosmic microwave background anisotropies
and big-bang nucleosynthesis in combination seem to favor an excess of radiation compatible with one family of sub-eV
sterile neutrinos~\cite{GonzalezGarcia:2010un,Hamann:2010bk,Giusarma:2011ex,Hou:2011ec}. On the other hand,
eV-mass sterile neutrinos are cosmologically viable only if additional ingredients are included since otherwise
sterile neutrinos would contribute too much hot dark matter~\cite{Hamann:2011ge} (see also \cite{Dodelson:2005tp}).

However, cosmological constraints during the BBN epoch have usually been derived under the assumption that the extra sterile neutrino
families were fully thermalised \cite{Hamann:2011ge}.
However, the validity of this assumption is not a priori clear and some preliminary studies \cite{Chu:2006ua,Melchiorri:2008gq}
already pointed toward this direction. It was shown in \cite{Abazajian:2004aj} that for plausible
values of the mass and mixing parameters, and initial lepton asymmetries not excluded by current observations there are cases where
little or no thermalisation occurs. For the charged fermions of the standard model the particle anti-particle asymmetry is known to be of
order $10^{-10}$. For neutrinos, however, no such bound exists, and the asymmetry can be many orders of magnitude larger without
violating observational constraints. In the standard model with no sterile states the upper bound on the neutrino chemical
potential is of order $\mu/T \lesssim {\rm few} \times 10^{-2}$
\cite{Kang:1991xa,Simha:2008mt,Castorina:2012md,Dolgov:2002ab,Wong:2002fa,Abazajian:2002qx} and while no exact bound has been derived in models with sterile neutrinos, we expect that the upper bound is of the same order of magnitude.

The purpose of this paper is to quantitatively derive the amount of thermalisation as a function of neutrino parameters
(mass, mixing, and initial lepton asymmetry).
We solve the full quantum kinetic equations in the 1 active+1 sterile approximation, calculate the effective
number of thermalised species just before BBN starts (at $T \simeq 1$~MeV)
and define under which conditions the thermalisation hypothesis holds. The assumption of $(1+1)$ families to evaluate the
thermalisation degree is justified for small lepton asymmetries since the resonances in the active sector are decoupled from the
conversions occurring in the active-sterile sector due to the larger mass difference. However, for large asymmetries active-sterile
conversion is delayed and can occur simultaneously with active-active conversion. While this does not qualitatively change
the overall picture there are some issues which we will return to in Section 3.

In our study, we calculate the number of thermalised extra families for the allowed mass-mixing parameter space for different
initial lepton asymmetries.  In Section~2 we introduce the adopted formalism and the quantum kinetic equations.
In Section~3, we present our results for initial null and large ($L= 10^{-2}$) lepton asymmetry.
 Conclusions and perspectives are presented in Section~4.

\section{Equations of motion}
\label{sec:eom}
In this section, we introduce the quantum kinetic equations (QKEs) governing the evolution of neutrinos in the early universe
\cite{Stodolsky:1986dx,Enqvist:1991qj,McKellar:1992ja,Sigl:1992fn,Boyanovsky:2007zz,Barbieri:1990vx}.
We adopt a mapping of the Bloch vectors
in terms of new vectors related to the active and sterile species grouping large and small dynamical variables.

\subsection{Quantum Kinetic Equations}
\label{sec:qke}
We consider oscillations of one active flavour $\nu_a$ (with $a=e$ or $\mu,\tau$) with a sterile neutrino
state $\nu_s$. Denoting with $\theta_s$ the mixing angle in vacuum and with $\nu_1$ and $\nu_2$ the
two mass eigenstates, separated by the mass difference $\delta m^2_s$, we have:
\begin{eqnarray}
\nu_a= \cos \theta_s \nu_1 - \sin \theta_s \nu_2\ , \\
\nu_s= \sin \theta_s \nu_1 + \cos \theta_s \nu_2\ .
\end{eqnarray}
In what follows we will refer $\delta m_s^2 >0$ as the normal hierarchy scenario (NH) and $\delta m_s^2<0$ as the inverted hierarchy
scenario (IH). Structure formation data strongly disfavour models with a total thermalised neutrino mass (the sum of all fully thermalised mass states)
in excess of 0.5-1 eV. Given that all the active states are fully thermalised this disfavours the inverted hierarchy for sterile masses above 0.2-0.3 eV. However, for masses below this the inverted hierarchy is not disfavoured and for completeness we study the same mass and mixing parameter space for both
NH and IH.

In order to describe the evolution of sterile neutrinos in the early universe, we use the density matrix
formalism and we express the density matrix associated with each momentum $p$ in terms
of the Bloch vector  components $(P_0,\mathbf{P})=(P_0, P_x,P_y,P_z)$~\cite{Enqvist:1991qj,Stodolsky:1986dx,Sigl:1992fn},
\begin{eqnarray}
\rho = \frac{1}{2} f_0 (P_0+ {\bf P} \cdot {\bf\sigma}) \; , \qquad
\overline{\rho} = \frac{1}{2} f_0 (\overline P_0+ {\bf \overline P} \cdot {\bf \sigma}) \;,
\end{eqnarray}
where $\mathbf{\sigma}$ are the Pauli matrices and $f_0 = 1/(1+e^{p/T})$ is the Fermi-Dirac distribution function with no chemical potential.
The neutrino kinetic equations in terms of the components of the Bloch vectors
for each momentum mode are:
\begin{eqnarray}
\label{qkep}
\dot{\mathbf{P}} &=& \mathbf{V} \times \mathbf{P}-D (P_x \mathbf{x}+P_y \mathbf{y})+\dot{P}_0 \mathbf{z}\ ,\\
\label{qkep0}
\dot{P}_0 &=&\Gamma \left[\frac{f_{\rm eq}}{f_0}-\frac{1}{2} (P_0+P_z)\right]
\end{eqnarray}
where the dot denotes the time derivative ($\di_t = \partial_t - H p \partial_p$, with $H$ the
Hubble parameter) and $f_{\rm eq}= 1/(1 + e^{(p-\mu)/T})$.

Defining the comoving momentum $x=p/T$, the vector $\mathbf{V}$ has the following components
\begin{eqnarray}
    V_x &=& \frac{\delta m_s^2}{2 xT} \sin 2 \theta_s \;,\\
V_y &=& 0\ ,\\
    V_z  &=& V_0  + V_1  + V_L.
\label{vzeta}
\end{eqnarray}
and
\begin{align}
  V_0		&= - \frac{\delta m_s^2}{2 x T} \cos 2 \theta_s, \\
  V_1^{(a)} 	&= - \frac{7 \pi^2}{45\sqrt{2}} \frac{G_F}{M_Z^2} x T^5
						\left[ n_{\nu_a} + n_{\bar{\nu}_a } \right] g_a\\
  V_L		&=  \frac{2 \sqrt{2}\zeta (3)}{\pi^2} G_F T^3 L^{(a)}. \label{eq:VL}
\end{align}
 Here, $g_{\mu,\tau}=1$ for $\nu_{\mu,\tau}$--$\nu_s$ mixing, $g_e=1+4\sec^2\theta_W/(n_{\nu_e} + n_{\bar{\nu}_e})$ for $\nu_e$--$\nu_s$ mixing and  $\theta_W$ is the Weinberg angle. The dimensionless number densities $n_{\nu_{a},(\bar{\nu}_a)}$ are the equilibrium active neutrino (antineutrino) densities normalised to unity in thermal equilibrium. The effective neutrino asymmetries $L^{(a)}$ are defined by
\begin{eqnarray}
    L^{(e)}    &=&  \left( \frac{1}{2} + 2 \sin^2 \theta_W \right) L_e
                  + \left(\frac{1}{2} - 2 \sin^2 \theta_W \right) L_p
                  - \frac{1}{2} L_n + 2 L_{\nu_e}
                  + L_{\nu_{\mu}} + L_{\nu_\tau}\ , \\
    L^{(\mu)}  &=&  L^{(e)} - L_e -L_{\nu_e} + L_{\nu_{\mu}}\ , \\
    L^{(\tau)} &=&  L^{(e)} - L_e -L_{\nu_e} + L_{\nu_{\tau}}\ ,
\end{eqnarray}
where $L_f \equiv (n_f - n_{\bar f})N_f/N_\gamma$ with $N_f$ ($N_\gamma$) the integrated active (photon) number density in
thermal equilibrium.
The potential $V_L$, defined as in Eq.~(\ref{eq:VL}), is the leading order contribution to $V_z$. The $V_1$ term
is the finite temperature correction and for example in the case of $\nu_e$--$\nu_s$ mixing
it includes  coherent interactions of  $\nu_e$ with the medium through which it propagates.
The condition for a matter induced resonance to occur is $V_z = 0$, and because $V_z$ depends on $L^{(a)}$ any non-zero lepton asymmetry
can have dramatic consequences for oscillation driven active-sterile neutrino conversion. In Appendix A we discuss the location of resonances in detail
for all possible values of mass, mixing, and lepton asymmetry.

A detailed derivation of the quantum kinetic equations is presented in \cite{Enqvist:1990ad,McKellar:1992ja}. Here we choose to adopt minimal assumptions on the collision terms. In particular, the term $D$ is the damping term, quantifying the loss of quantum coherence due to
$\nu_a$ collisions with the background medium. For example, considering $\nu_e$,
the elastic contribution should come from the elastic scattering of $\nu_e$ with
$e^-$ and $e^+$ and with the other active flavours $\nu_a$ and $\bar{\nu}_a$. The inelastic
contribution comes from the scattering of $\nu_e$ with $\bar{\nu}_e$ (producing
$e^-$ and $e^+$ or $\nu_a$ and $\bar{\nu}_a$).   In terms of the Bloch vectors such
terms have the effect of suppressing the off-diagonal elements of the density matrix ($P_{x,y}$).
The effective potentials contributing to this term have been previously calculated~\cite{Enqvist:1990ek,Enqvist:1990ad,Notzold:1987ik} and
if thermal equilibrium is aasumed and the electron mass neglected, it is approximately half the corresponding scattering rate $\Gamma$ \cite{Stodolsky:1986dx,McKellar:1992ja,Bell:1998ds}
\begin{equation}
    D = \frac{1}{2} \Gamma\ . \label{eq:damp}
\end{equation}

The evolution of $P_0$ is determined by processes that deplete or enhance the
abundance of $\nu_a$ with the same momentum and its rate of change receives no contribution from coherent
$\nu_a$-$\nu_s$ oscillations.
The repopulation term $\Gamma (f_{\rm eq}/f_0-1/2 (P_0+P_z)$) is an approximation for the
correct elastic collision integral~\cite{Bell:1998ds} with
\begin{equation}
    \Gamma = C_a G_F^2 x T^5 \label{eq:repopulation}
\end{equation}
where $C_e \simeq 1.27$ and $C_{\mu,\tau} \simeq 0.92$~\cite{Enqvist:1991qj}.
Note that the term including the effective collision rate, $\Gamma$, is an approximation to the full momentum dependent scattering kernel which repopulates neutrinos from the background plasma.
The full expression has been derived in \cite{McKellar:1992ja}. In \cite{Bell:1998ds} it was proven that the general form of $D$ (and $\Gamma$) exactly reduces to Eqs.~(\ref{eq:damp},\ref{eq:repopulation})  for weakly interacting
species in thermal equilibrium with zero chemical potential, and that it is the zero order approximation for particles with non-null chemical potential.
The respective  equations of motion for anti-neutrinos can be found by substituting
$L^{(a)} = -L^{(a)}$ and $\mu = - \mu$ in the above equations.
In our treatment we have not included the rate equations for the electrons and positrons since
 we are assuming that all the species electromagnetically interacting  are kept in equilibrium.

\subsection{Mapping with the active and the sterile variables}
\label{sec:changevariables}

We can distinguish among large and small linear combinations of the dynamical variables in the
particle and antiparticle sector to simplify the numerical treatment. For each momentum
mode, we define for each component $i$ (with $i=0,x,y,z$) of the Bloch vector
\begin{equation}
P_i^{\pm}=P_i \pm \overline{P}_i\ .
\end{equation}
We also separate active ($a$) and sterile ($s$) sectors
\begin{eqnarray}
P_a^{\pm}&=&P_0^{\pm}+P_z^{\pm}=2\frac{\rho^{\pm}_{aa}}{f_0}\ ,\\
P_s^{\pm}&=& P_0^{\pm}-P_z^{\pm}=2\frac{\rho^{\pm}_{ss}}{f_0}\ .
\end{eqnarray}
Therefore, in terms of the new vectors Eqs.~(\ref{qkep}, \ref{qkep0}) become
\begin{eqnarray}
\label{pa}
\dot{P}_a^\pm & = &  V_x P_y^\pm + \Gamma  \left[2 f_{eq}^\pm/f_0
                      - P_a^\pm \right]\ ,\\
\label{ps}
\dot{P}_s^\pm & = &   -   V_x P_y^\pm\  ,\\
\label{px}
\dot{P}_x^\pm & = &   -  (V_0 + V_1) P_y^\pm  - V_L P_y^\mp - D P_x^\pm\ , \\
\label{py}
\dot{P}_y^\pm & = & (V_0 + V_1) P_x^\pm  + V_L P_x^\mp
                  - \frac{1}{2} V_x (P_a^\pm - P_s^\pm)  - D P_y^\pm\ ,
\end{eqnarray}
where we have defined $f_{\rm eq}^\pm  =  f_{\rm eq} (p,\mu) \pm f_{\rm eq} (p,-\mu)$.

The lepton number can be directly calculated  from the integral over the difference between the neutrino and the antineutrino distribution functions, i.e. $P_a^-$:
\begin{equation}
L^{(a)} = \frac{2}{8\zeta(3)}\int\limits_0^\infty \text{d} x x^2 \rho_{aa}^-
	= \frac{1}{8\zeta(3)}\int\limits_0^\infty \text{d} x x^2 f_0 P_a^-\ . \label{eq:L_from_integral}
\end{equation}

However, since the repopulation term is approximated by Eq.~\eqref{eq:repopulation} which does not explicitly conserve lepton number  we independently evolve $L^{(a)}$ as in~\cite{DiBari:1999vg} using an evolution equation where the repopulation term does not enter. Taking the time derivative of Eq.~\eqref{eq:L_from_integral} and ignoring the repopulation part of $\dot{P}_a^-$, the evolution equation for $L^{(a)}$ is
\begin{equation}
\label{Lparevol}
  \dot{L}^{(a)} = \frac{1}{8\zeta(3)} \int\limits_0^\infty \text{d}x x^2 f_0 V_x P_y^-\ .
\end{equation}

Note that, in kinetic equilibrium, $\mu$, or rather
the degeneracy parameter $\xi\equiv\mu/T$, is related to the lepton number $L^{(a)}$ through the integral over $f_{\rm eq}^-$ \cite{Kang:1991xa}
\begin{equation}
L^{(a)}_{\text{eq}} = \frac{1}{4\zeta(3)} \int_0^\infty \text{d}x\ x^2 \[ \frac{1}{1+e^{x-\xi}}-\frac{1}{1+e^{x+\xi}} \] =
\frac{1}{12\zeta(3)} \( \pi^2\xi + \xi^3 \).
\end{equation}
This is a third order equation, and using Chebyshev's cubic root, one can extract the corresponding and expression for $\xi$
valid for any $L^{(a)}$ using trigonometric functions:
\begin{equation}
\xi = \frac{-2\pi}{\sqrt{3}} \sinh\(\frac{1}{3} \text{arcsinh}\[-\frac{18 \sqrt{3}\zeta(3)}{\pi^3} L^{(a)} \] \).
\end{equation}

In order to numerically solve the QKEs, we define the momentum grid in comoving coordinates ($x=p/T$). Therefore the grid becomes stationary and the partial differential equations become ordinary differential equations coupled through integrated quantities only. Using
the temperature $T$ as the evolution parameter, time derivatives, $\text{d}_t$, are replaced by $\rightarrow -HT \partial_T$ in the above equations, provided that the time derivative of the effective number of degrees of freedom can be ignored.

\section{Results: thermalised sterile species}
 The fraction of sterile thermalised species is defined as
 \begin{equation}
 \delta N_\text{eff,s} = \frac{\int \text{d} x x^3  f_0 P_s^+}{4 \int \text{d} x x^3 f_0}\ .
 \end{equation}
However, the total amount of radiation is given by the sum of active and sterile energy densities
 \begin{equation}
 \delta N_\text{eff} = \frac{\int \text{d} x x^3  f_0 \(P_s^+ + P_a^+-4\)}{4 \int \text{d} x x^3 f_0}\ .
 \label{Neff}
 \end{equation}
Note that when the active state is in thermal equilibrium ($P_a^+=4$), $\delta N_\text{eff,s} =  \delta N_\text{eff}$. When $L^{(a)}$ is large, the sterile sector may be populated so late that the active sector does not have time to repopulate before it decouples.
In this section, we discuss the fraction of thermalised species for initial $L^{(a)}=0$ and $L^{(a)}=10^{-2}$
and for a range of ($\delta m^2_s$, $\sin^2 2\theta_s$).

In terms of late-time cosmological constraints on light neutrinos both $\delta N_\text{eff,s}$ and $\delta N_\text{eff}$ can be relevant quantities. Models with a modified light neutrino sector are most often parametrised in terms of the neutrino mass, $m_\nu$, and $N_{\rm eff}$ in such a way that $N_{\rm eff}$ neutrino species all share the same common mass $m_\nu$ (i.e.\ it is assumed that the mass spectrum is degenerate). However, in models with a single sterile state one instead has either $\delta N_\text{eff,s}$ steriles with mass $m_s$ and $3.046+\delta N_\text{eff}-\delta N_\text{eff,s}$ massless active states (NH) or
$3.046+\delta N_\text{eff}-\delta N_\text{eff,s}$ massive active state with degenerate mass and $\delta N_\text{eff,s}$ massless sterile states (IH).
These two cases are different when it comes to structure formation and should in principle be treated separately (see e.g.\ \cite{Hannestad:2006mi} for a discussion about this point).
Since the goal of this paper is to calculate $\delta N_\text{eff}$, not to provide quantitative constraints on specific models, we simply use $\delta N_\text{eff}$ from this point on.

\subsection{Numerical solution of the quantum kinetic equations}
Solving the quantum kinetic equations numerically is non-trivial task. The number of differential equations are roughly $8N$ where $N$ is the number of momentum bins, and since the resonances can be very narrow we need a few hundred points to obtain good precision. There are many vastly separated time-scales involved, so the problem is stiff, and once $L$ changes, the system becomes extremely non-linear. We used two different solvers, one based on the numerical differentiation formulae of order $1-5$ ({\tt ndf15}) due to Shampine~\cite{Shampine1997}, and one based on the fifth order implicit Runge-Kutta method {\tt RADAU5} due to Hairer and Wanner~\cite{hairer1993solving}. If the maximum order of the first method is reduced to two, both solvers are L-stable, and thus excellent for stiff problems. Because of the large number of equations and the sparsity of the Jacobian we must use sparse matrix methods for the linear algebra operations needed in both solvers. For this purpose, we are employing a small sparse matrix package based on~\cite{Davis2006}.

To sample the momentum-space in an optimal way we are mapping the $x$-interval $[x_\text{min}; x_\text{max}]$ to a $u$-interval $[0; 1]$ by
\begin{align}
u(x) &= \frac{x-x_\text{min}}{x_\text{max}-x_\text{min}} \times \frac{x_\text{max}+x_\text{ext}}{x + x_\text{ext}},
\end{align}
where $x_\text{ext}$ is the extremal point of some moment of the Fermi-Dirac distribution. We chose the values $x_\text{min}=10^{-4}$, $x_\text{ext}=3.1$ and $x_\text{max}=100$, and then sampled $u$ uniformly. This is the same mapping employed in~\cite{Kainulainen:2001cb}, but they go one step further and introduce an adaptive grid that follows the resonances. This is not necessary for this project since our mixing angles are comparably larger,  and we are not looking at chaotic amplification of an initially small value of $L$.

We evolved the system from an initial temperature of $60$~MeV to a final temperature of $1$~MeV for the following grid of masses and mixing angles:
\begin{subequations}
\begin{align}
\label{range}
10^{-3}\ \text{eV}^2 \le \delta m^2_s \le 10\ \text{eV}^2 \text{ and } 10^{-4} \le \sin^2 2 \theta_s \le 10^{-1} \text{ for } \qquad L^{(a)}&=0\ , \\
\label{zoomrange}
10^{-1}\ \text{eV}^2 \le \delta m^2_s \le 10\ \text{eV}^2 \text{ and } 10^{-3.3} \le \sin^2 2 \theta_s \le 10^{-1} \text{ for }  \qquad L^{(a)}&=10^{-2}\ .
\end{align}
\end{subequations}
We ran the complete grids for different number of momentum bins, different accuracy parameters and both differential equation solvers with no noticeable difference.
\subsection{Sterile neutrino production for zero lepton asymmetry}\label{sec:smallasy}

The simplest case, and the one most often studied in the literature, is the one where the lepton asymmetry is zero.
For $L^{(a)}=0$, the evolution of $P_i^{+}$ is decoupled from  $P_i^{-}$ [see Eqs.~(\ref{pa},\ref{py})] and the asymmetry remains zero for the whole evolution (as can be seen from Eq.~(\ref{Lparevol})).

From Eqs.~(\ref{eq:xres},\ref{eq:Fsymbol}) in Appendix A it can be seen that there is either no resonance (NH) or that the resonances are identical
for neutrinos and anti-neutrinos (IH). As it is well known, in IH the resonance propagates to higher values of $x$ as the universe expands and eventually covers
the entire momentum distribution of neutrinos.

\begin{figure}[t]
\centering
\includegraphics[angle=270, width=0.7\textwidth]{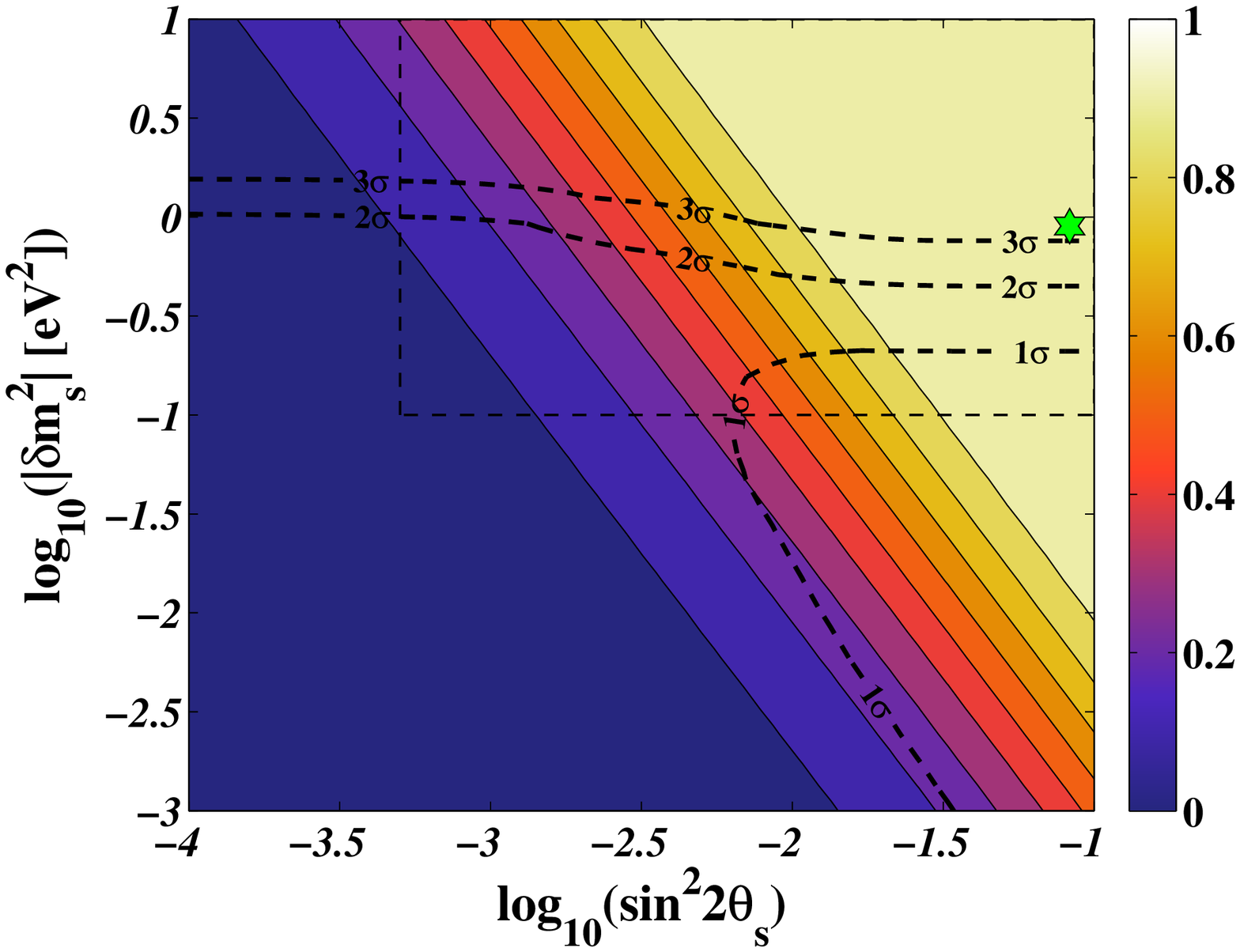}
\includegraphics[angle=270,width=0.7\textwidth]{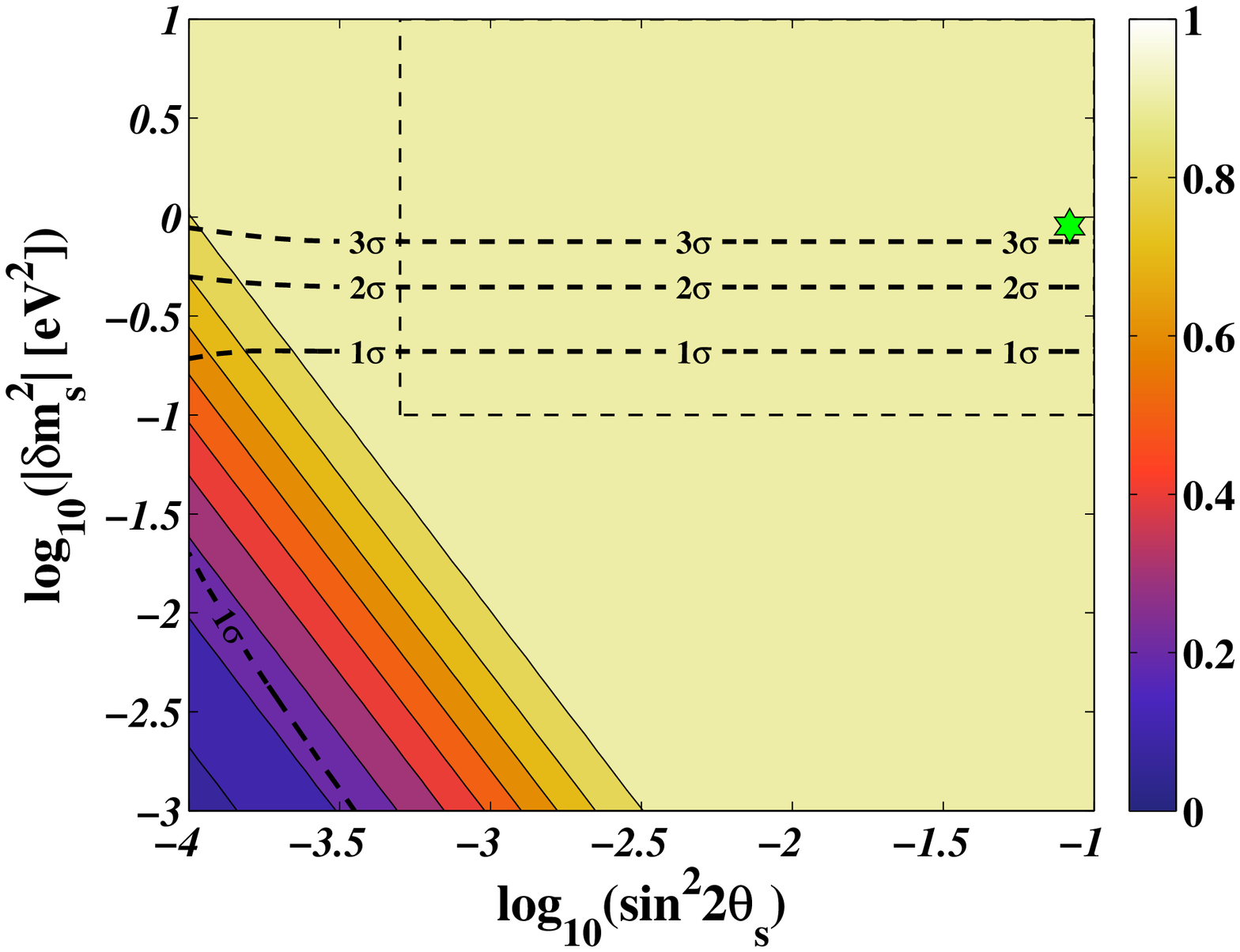}
 \caption{Iso-$\delta N_{\rm eff}$ contours in the $\sin^2 2\theta_s-\delta m_s^2$ plane for $L^{(\mu)}=0$ and $\delta m^2_s > 0$ (top panel) and $\delta m^2_s < 0$ (bottom panel).
 The green hexagon denotes the $\nu_s$ best-fit mixing parameters as in the $3+1$ global fit in~\cite{Giunti:2011cp}: $(\delta m^2_s,\sin^2 2\theta_s)=(0.9\ \text{eV}^2,0.089)$.
 The $1-2-3 \sigma$ contours denote the CMB+LSS allowed regions for $\nu_s$ with sub-eV mass  as in~\cite{Hamann:2010bk}.
 In order to facilitate the comparison with the results presented in Sec~\ref{sec:largeleptona}, a dashed rectangle denotes the parameter-space described
 by~\eqref{zoomrange}.
 \label{fig:NHIH_zero}}
\end{figure}

 In Fig.~\ref{fig:NHIH_zero} we show the fraction of thermalised neutrinos, $\delta N_{\rm eff}$, for the range of mixing parameters given in
 Eq.~(\ref{range}) with initial asymmetry $L^{(\mu)} = 0$. The top panel shows the normal hierarchy, $\delta m_s^2 > 0$, and the bottom panel the inverted hierarchy, $\delta m_s^2 < 0$.
 The smaller parameter space described by~\eqref{zoomrange} is denoted with a dashed rectangle to facilitate comparison
 with the results presented in Sec~\ref{sec:largeleptona}.

We mark with a green hexagon the best fit point of the $3+1$ global analysis presented in~\cite{Giunti:2011cp}, obtained
from a  joint analysis of Solar, reactor, and short-baseline neutrino oscillation data
$(\delta m^2_s, \sin^2 2 \theta_s) = (0.9~{\rm eV}^2, 0.089)$. For that point $\delta N_\text{eff} = 1$ in both hierarchies, i.e.\
complete thermalization occurs.
In addition we show the parameter range preferred by CMB and large scale structure (LSS) data.
The $1-2-3 \sigma$ contours have been obtained interpolating the likelihood function obtained in~\cite{Hamann:2010bk}
for each fixed $\delta m^2_s$ and $N_\text{eff}$.
In both cases the lower left corners of parameter space where little thermalization occurs are disfavoured because of the
CMB+LSS preference for extra energy density.

\begin{figure}[t]
\centering
\includegraphics[angle=270, width=0.65\textwidth]{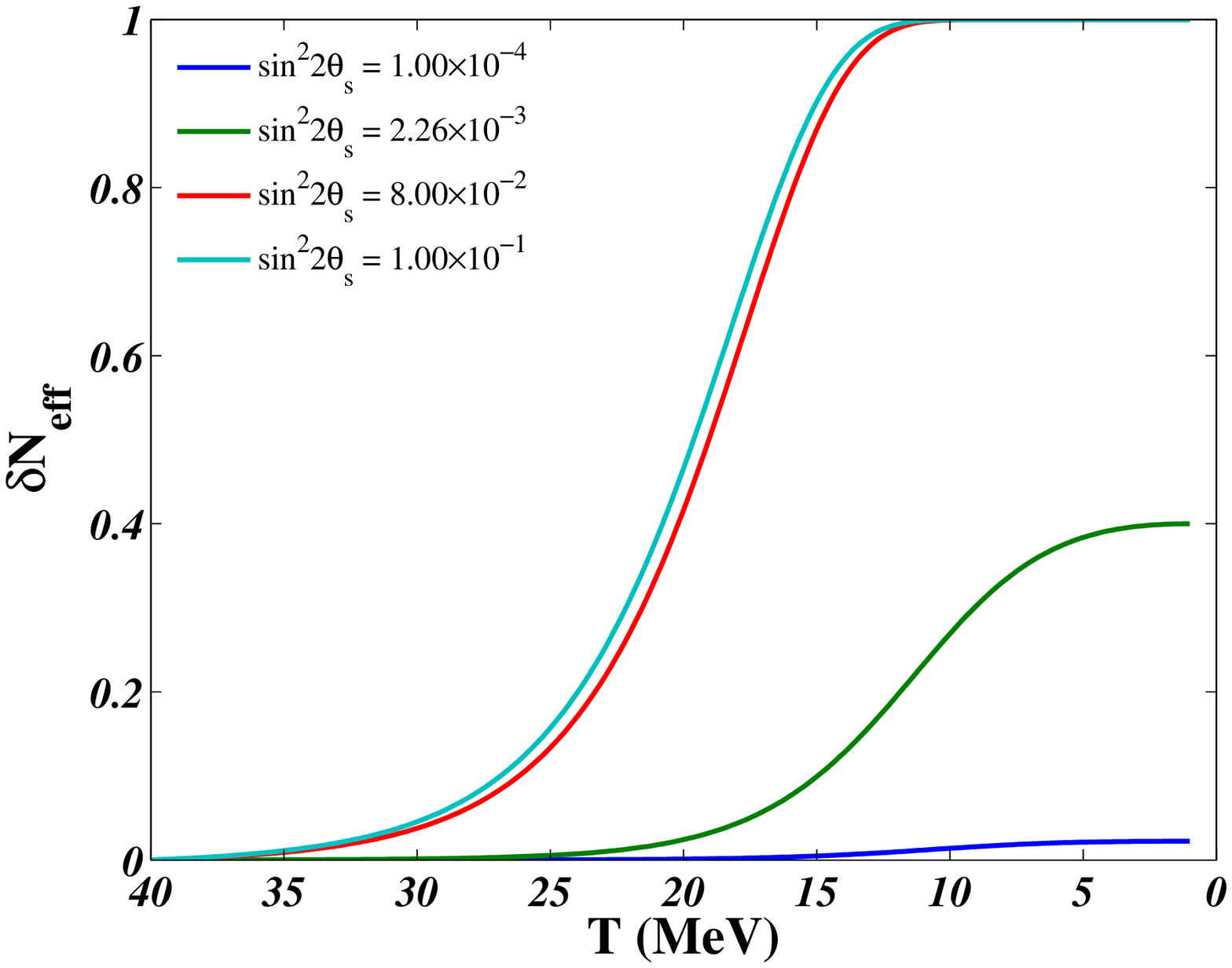}
\includegraphics[angle=270, width=0.65\textwidth]{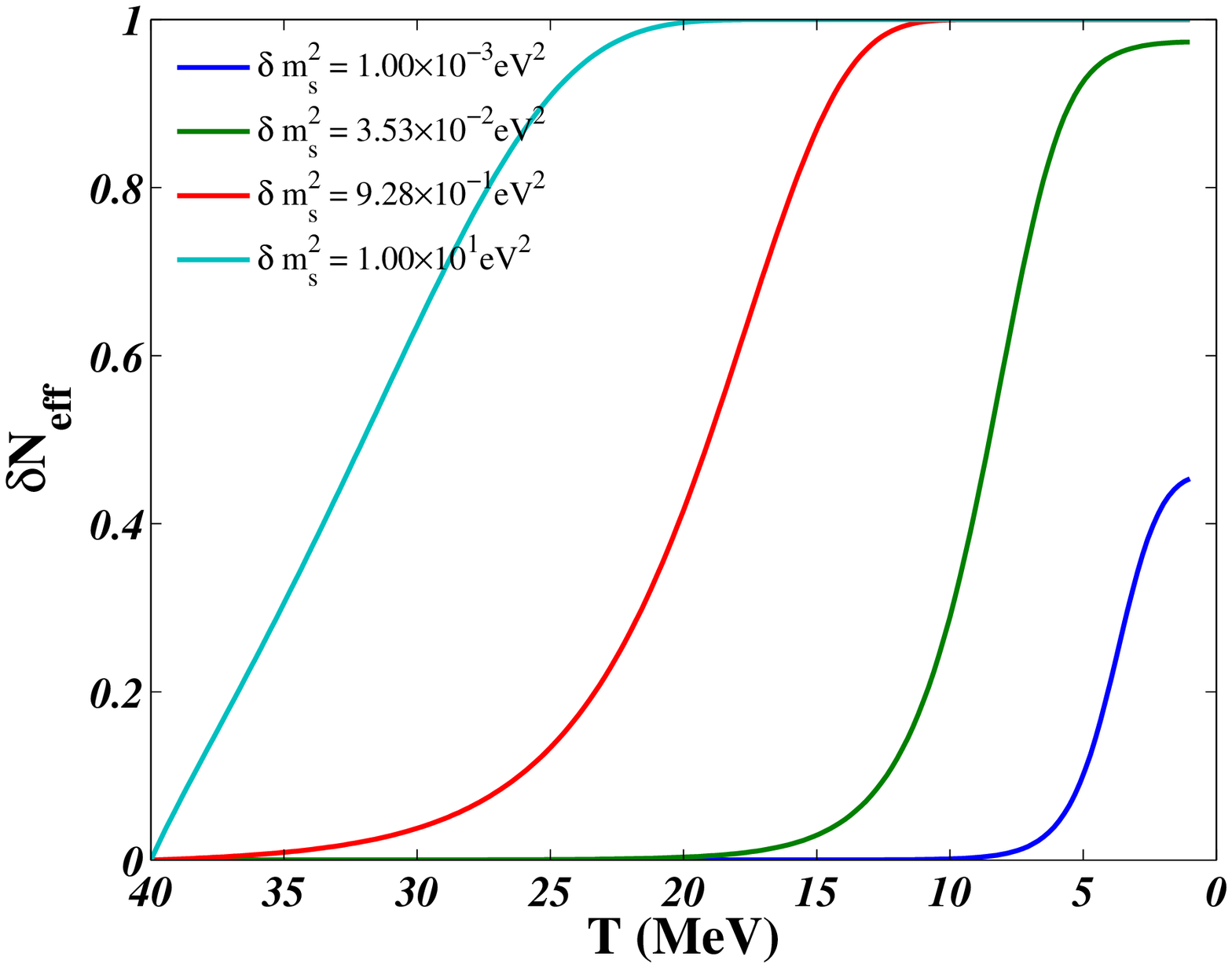}
 \caption{Top panel: $\delta N_{\rm eff}$ as a function of the temperature for four different mixing angles ($\sin^2\ 2 \theta_s = 10^{-4}, 2 \times 10^{-3}, 5 \times 10^{-2}, 10^{-1}$) and fixed mass difference ($\delta m^2_s = 0.93\ \text{eV}^2$). Bottom panel: $\delta N_{\rm eff}$ as a function of the temperature for four different mass differences ($\delta m^2_s = 10^{-3}, 3.5 \times 10^{-2}, 9.3 \times 10^{-1}, 10$~eV$^2$) and fixed mixing angle ($\sin^2 2\theta_s = 0.051$).  Thermalisation begins earlier and is more effective for larger mass differences and for larger mixing angles.
} \label{fig:dm2sin2fixed}
\end{figure}

It is also of interest to see how the thermalization proceeds as a function of temperature. In Fig.~\ref{fig:dm2sin2fixed} we show
the evolution of $\delta N_\text{eff}$ as a function of temperature for the NH scenario for a variety of different
$\delta m^2_s$ and $\sin^2 2 \theta_s$.
For the non-resonant NH, the thermalization rate of sterile neutrinos is approximately $\Gamma_s \sim \frac{1}{2} \sin^2 2\theta_s \Gamma$. The maximum
thermalisation rate occurs at a temperature of approximately $T_{\rm max} \sim 10 \, (\delta m^2_s)^{1/6} \,\, {\rm MeV}$ and the final $\delta N_\text{eff}$ depends only on $\sin^2 2\theta_m$ at that temperature (see \cite{Enqvist:1991qj} for a detailed discussion).
In the top panel of Fig.~\ref{fig:dm2sin2fixed} this behaviour can be seen. For very large vacuum mixing $\Gamma_s/H > 1$ already before $T_{\rm max}$ such that
complete thermalisation has occurred already before $T_{\rm max}$ reached. For smaller mixing $\Gamma_s/H$ never exceeds 1 and even though thermalisation proceeds fastest around $T_{\rm max}$ it is never fast enough to equilibrate the sterile states.

In the bottom panel the change in $T_{\rm max}$ as $\delta m^2_s$ varies is evident, and provided that $T_{\rm max}$ is higher than the active neutrino decoupling temperature the vacuum mixing in this case is large enough that complete thermalisation always occurs.
For the non-resonant case the end result is that isocontours of $\delta N_\text{eff}$ always lie at constant values of $\delta m^2_s \sin^4 2 \theta_s$, as can
be seen in the top panel of Fig.~\ref{fig:NHIH_zero}.

\begin{figure}[t]
\centering
\includegraphics[angle=270, width=0.45\textwidth]{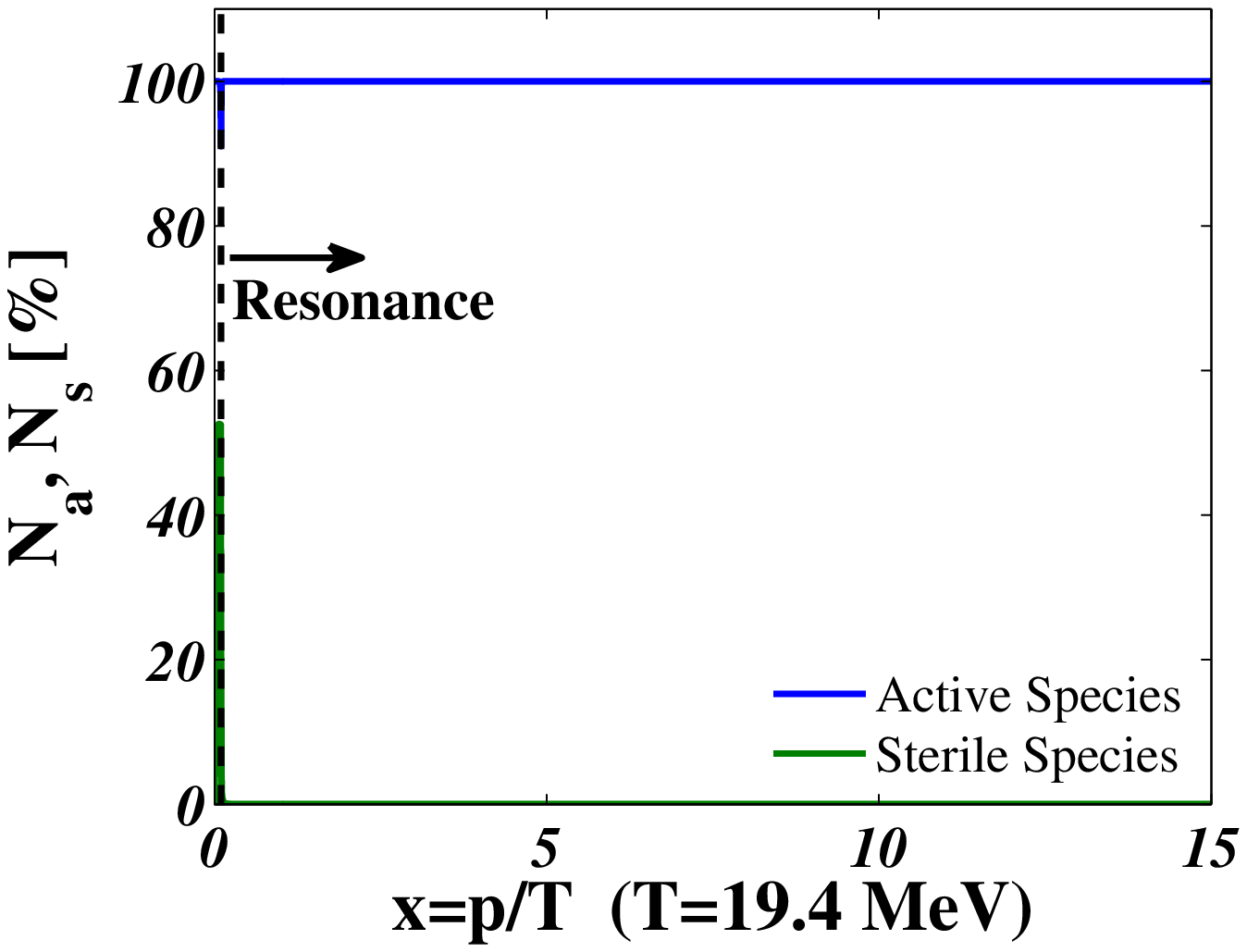}
\includegraphics[angle=270, width=0.45\textwidth]{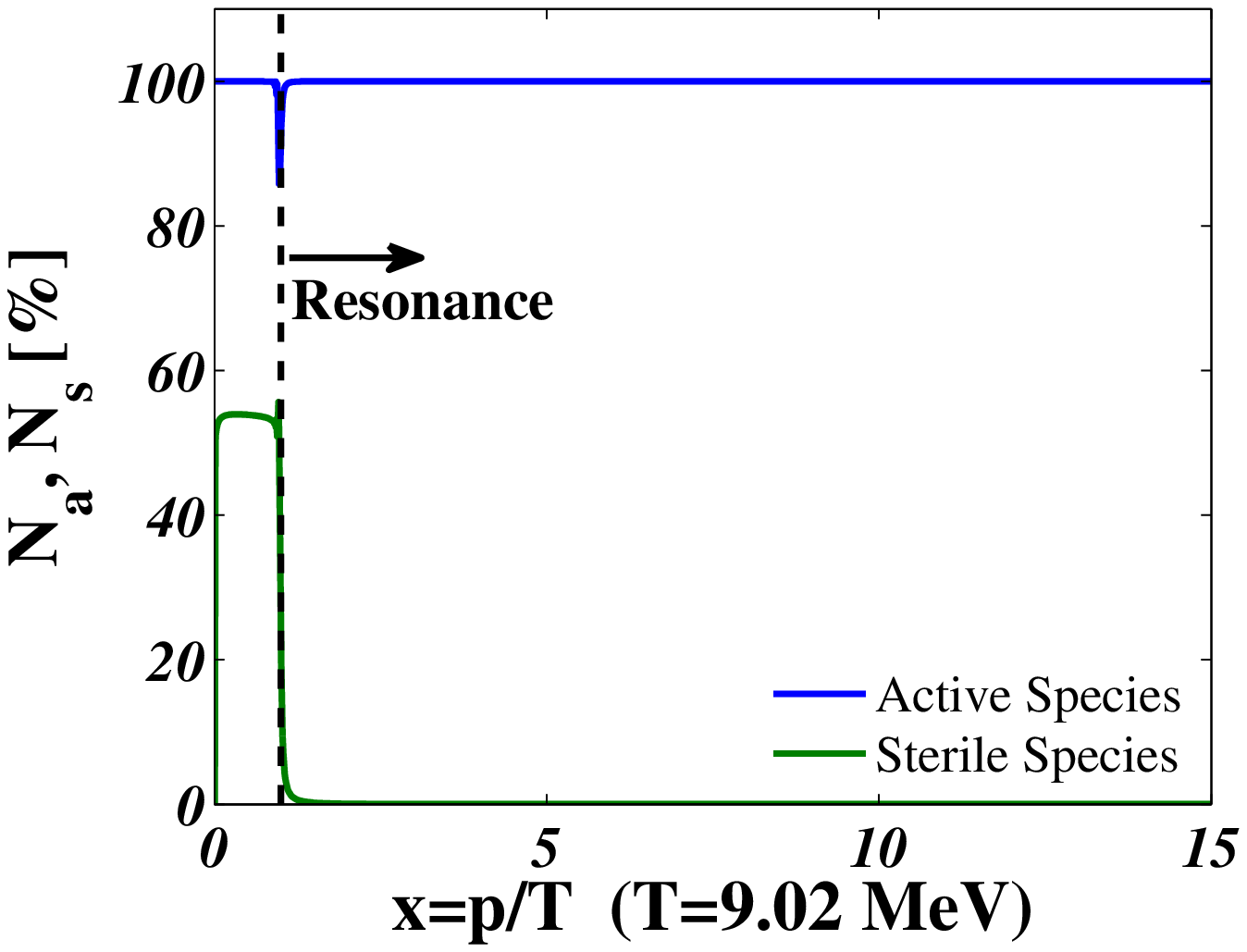}
\includegraphics[angle=270, width=0.45\textwidth]{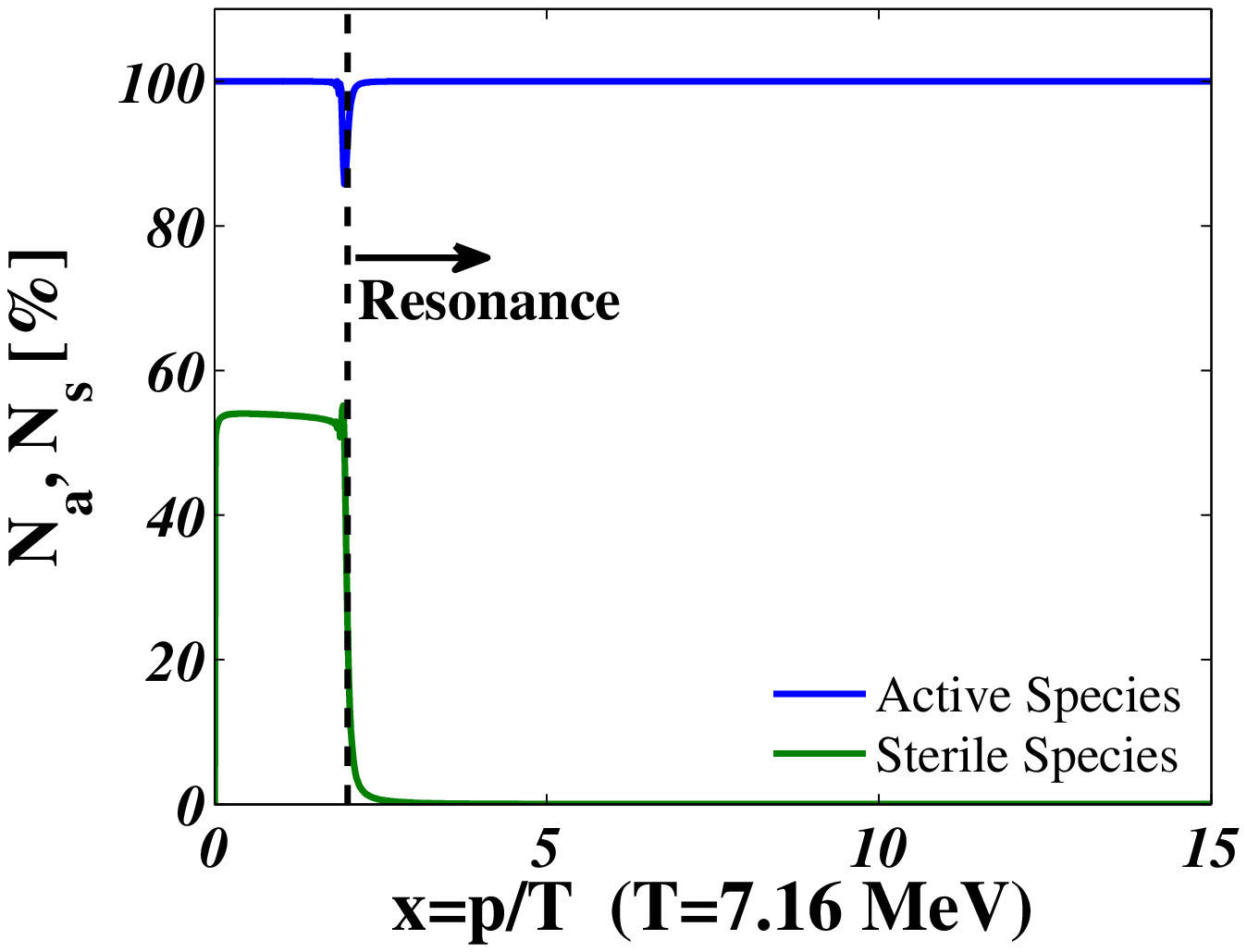}
\includegraphics[angle=270, width=0.45\textwidth]{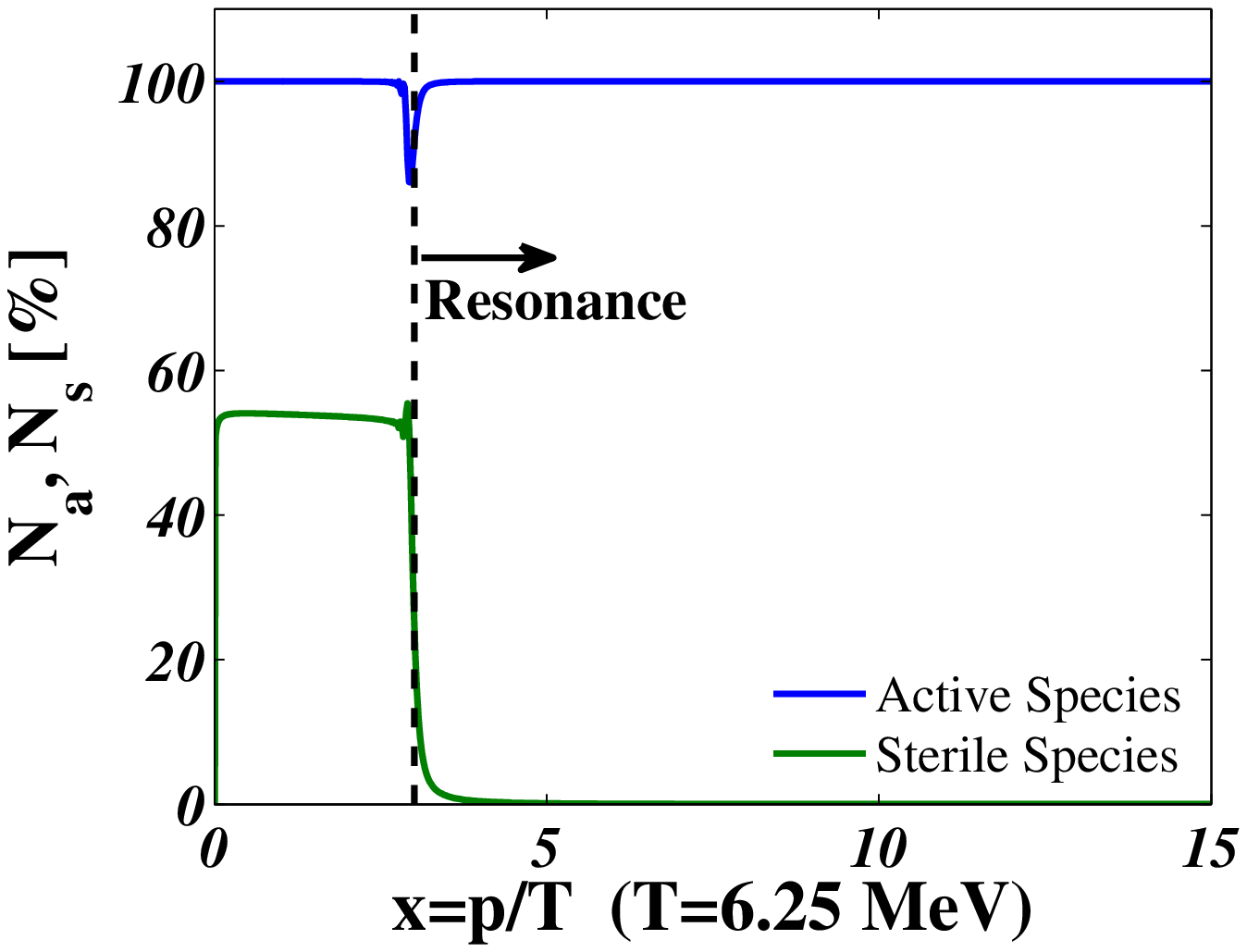}
\includegraphics[angle=270, width=0.45\textwidth]{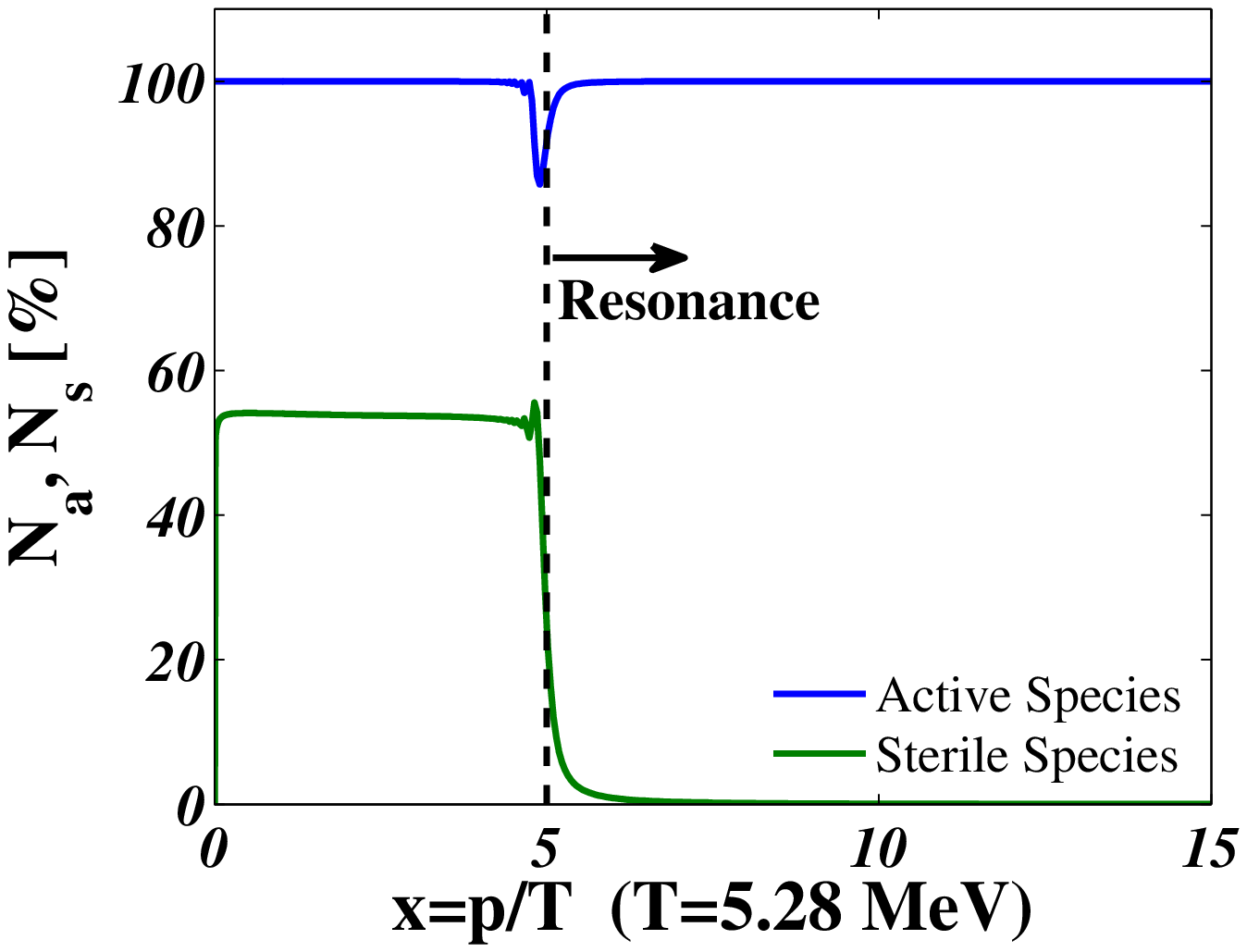}
\includegraphics[angle=270, width=0.45\textwidth]{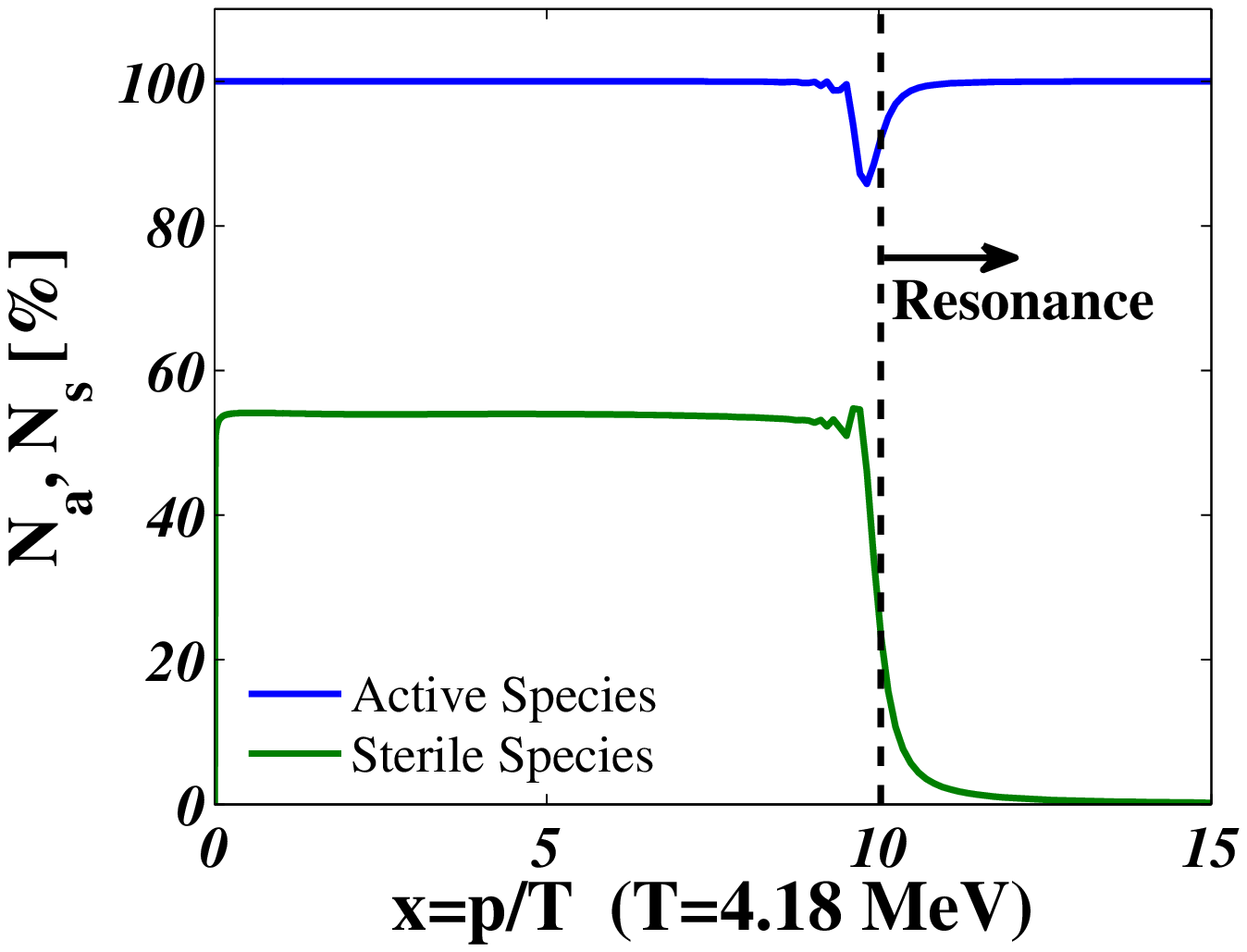}
 \caption{Temperature evolution of active and sterile neutrino distributions for the resonant case $(\delta m^2_s,\sin^2 \theta_s)=(-3.3\ \text{eV}^2,6 \times 10^{-4})$ and $L^{(\mu)}=0$. \label{fig:cartoon}}
\end{figure}

In the inverted hierarchy the resonance conditions are always satisfied. Therefore, we expect full thermalization for a larger region of the mass-mixing parameters than in NH, as confirmed in Fig.~\ref{fig:NHIH_zero}. In this case, thermalisation may proceed through resonant conversions alone.
For illustration, we choose the point of Fig.~\ref{fig:NHIH_zero} with  $(\delta m^2_s,\sin^2 \theta_s)=(-3.3\ \text{eV}^2,6 \times 10^{-4})$ for which $\delta N_{\rm eff}=0.55$ and we show the  percentage of active ($N_a$) and sterile ($N_s$) neutrinos as a function of $x$ for different $T$  in Fig.~\ref{fig:cartoon}. The thermalisation is not complete and it is  nearly instantaneous as the resonance moves through the momentum
spectrum and the resulting dip in the active sector is quickly repopulated from the background.
%

We have presented results for $L^{(\mu)}=0$ only, but the case of $L^{(e)}=0$ shows exactly the same trend as in Fig.~\ref{fig:NHIH_zero}. However, the region with  $\delta N_\text{eff} = 1$ is slightly smaller than the one shown in Fig.~\ref{fig:NHIH_zero}. This is due to the fact that $\nu_e$'s have a larger potential than $\nu_{\mu,\tau}$ (because of the charged current interaction contribution) and therefore resonances occur at slight lower temperatures.

\subsection{The case of large initial lepton asymmetry}
\label{sec:largeleptona}

\begin{figure}[t]
\centering
\includegraphics[angle=270, width=0.7\textwidth]{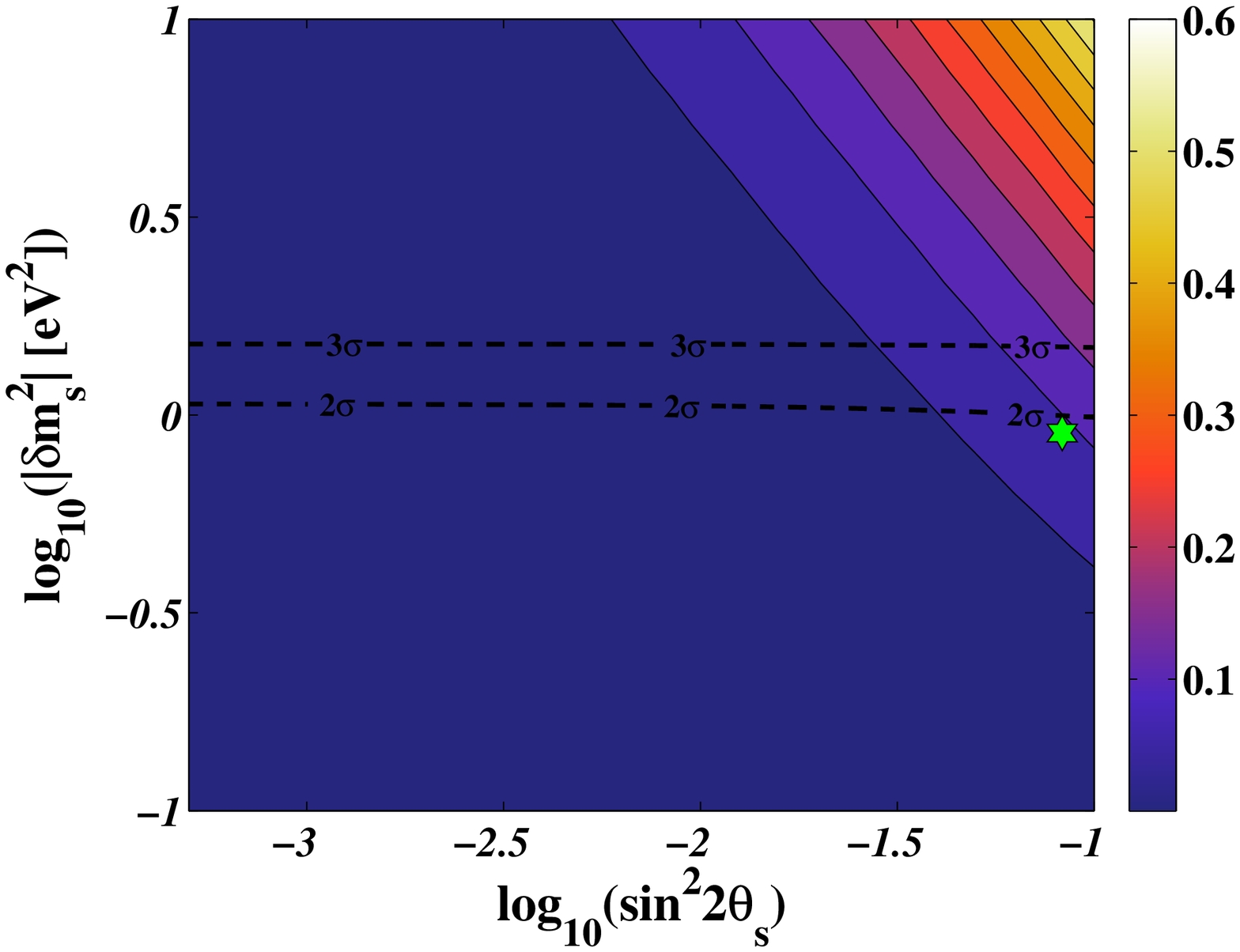}
\includegraphics[angle=270, width=0.7\textwidth]{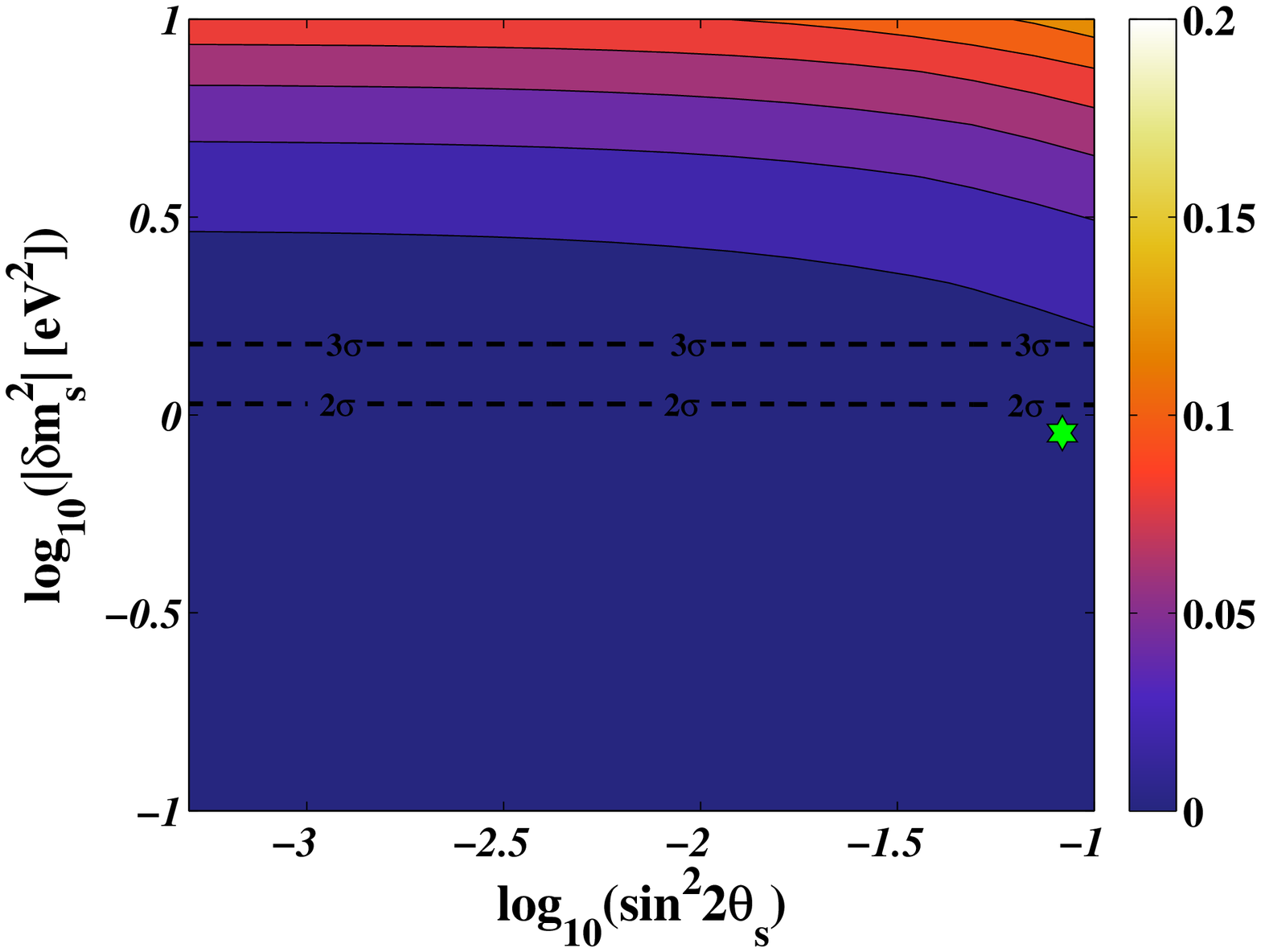}
 \caption{Iso-$\delta N_{\rm eff}$ contours in the $\sin^2 2\theta_s-\delta m^2_s$ plane for $L^{(\mu)}=10^{-2}$ and
 $\delta m^2_s > 0$ (top panel) and $\delta m^2_s < 0$ (bottom panel), as in Fig.~\ref{fig:NHIH_zero}. \label{fig:NHIH_1em2}}
\end{figure}

We now discuss the thermalisation degree for initial large lepton asymmetry. In principle, one would expect a lepton asymmetry of the same order of magnitude as the baryon asymmetry ($\eta \simeq 10^{-10}$). However, since neutrinos are neutral particles, $L^{(a)}=10^{-2}-10^{-1}$ is not presently excluded~\cite{Dolgov:2002ab,Pastor:2008ti,Castorina:2012md} by the requirement of charge neutrality. A large lepton asymmetry is responsible for blocking the active-sterile flavor conversions by an in-medium suppression of the mixing angle; therefore it has been invoked as a means of significantly reducing the sterile abundance~\cite{Chu:2006ua}. A large lepton number can be generated by e.g.\ an Affleck-Dine mechanism~\cite{Kawasaki:2002hq}
or other models that are able to produce large lepton asymmetries and small baryonic ones~\cite{Harvey:1981cu,Dolgov:2002wy}.
Another interesting possibility is to grow the lepton asymmetry from some initial $L^{(a)}\sim\mathcal{O}(10^{-10})$ using active-sterile oscillations~\cite{Foot:1995qk,Barbieri:1989ti,Barbieri:1990vx}. Solving the QKE's in IH and with an initially small but non-zero lepton number,  our preliminary results point toward a final lepton number varying between $10^{-5}$ and $10^{-2}$ depending on the mixing parameters.
For illustrative purposes, we choose to adopt $L^{(a)}=10^{-2}$.

Figure~\ref{fig:NHIH_1em2} shows the $\delta N_{\rm eff}$ contour plot for $L^{(\mu)}=10^{-2}$ and $\delta m^2_s > 0$ (top panel) and $\delta m^2_s < 0$ (bottom panel). The region with full thermalisation is now much smaller than in Fig.~\ref{fig:NHIH_zero}.

As we discuss in detail in Appendix A, a large value of $L^{(a)}$ confines the resonances to very small or large values of $x$, far away from the maximum of the active neutrino momentum distribution (see also \cite{Shi:1998km}).
Only at relatively low temperature does the resonance begin to move through the momentum distribution. What happens next is qualitatively very different
for normal and inverted hierarchy. For NH the lepton asymmetry decreases as the resonance moves. This causes a run-away effect because as $L^{(a)}$ decreases the resonance moves faster, causing a faster decrease in $L^{(a)}$. When $L^{(a)}$ becomes less than approximately $10^{-5}$ (see Eq.~\eqref{eq:A}), the resonance disappears and the remaining evolution after this point is equivalent to the $L^{(a)}=0$ NH case. For sufficiently large $\delta m^2_s$ and $\sin^2 2 \theta_s$ the non-resonant production after the resonance disappears can be significant. However, the required mass difference and mixing to obtain the same degree of thermalisation are much larger than in the $L^{(a)}=0$ case.

The rapid depletion of $L^{(a)}$ in NH causes the numerical solution to continue after this point with a very small time step. No further resonant production will occur after this point, but as we discuss above, some non-resonant thermalisation has yet to happen at this stage. To circumvent this problem, we stop the code when $L^{(a)}$ becomes very close to zero and restart it again with $L^{(a)}=0$ using the static approximation discussed in Appendix B.

For IH the lepton asymmetry increases when the resonance moves, causing it to move slower and effectively blocking population of the sterile state until very late. For the range of mixing parameters studied here production of sterile neutrinos is effectively blocked until after the active species decouples, leading to a very small $\delta N_\text{eff}$. In Appendix A we give equations for the position of the resonances for finite $L^{(a)}$ along with useful approximations valid in different limits.

For $\nu_s$ mixing parameters as in~\cite{Giunti:2011cp}, $\delta N_\text{eff} \sim 0$ in IH and $\delta N_\text{eff} = 0.05$ in NH.
Constraints from BBN, CMB, and LSS have usually assumed a fully thermalised sterile state, but as also mentioned in \cite{Hamann:2010bk}
a finite lepton asymmetry can effectively block thermalisation and make this assumption invalid. In that case an eV sterile neutrino
will not be in conflict with the cosmological neutrino mass bound, but of course the extra energy density preferred by CMB and LSS will then not
be associable with the light sterile neutrino.

We finally note that since we have solved the quantum kinetic equations using the 1 sterile + 1 active approximation, only one lepton asymmetry is relevant
in our equations (either $e$ or $\mu$). However, in the real 3+1 scenario there will be 3 separate flavour asymmetries and active-active oscillations will
lead to some degree of equilibration between these asymmetries. While there will be some quantitative differences between our 1+1 treatment and the full 3+1 scenario we do expect the same qualitative behaviour, i.e.\ a blocking of thermalisation due to confinement of the active-sterile resonances.

\section{Conclusions}
\label{sec:conclusions}

Recent cosmological data seem to favor an excess of radiation beyond three neutrino families
and photons, and light sterile neutrinos are possible candidates.  The upcoming measurement
of $\delta N_\text{eff}$ by Planck will confirm  or rule out the existence of such
extra radiation with high precision~\cite{Perotto:2006rj,Hamann:2007sb}.

Light sterile neutrinos could thermalise prior to neutrino decoupling, contributing to the relativistic energy density
in the early universe. Present data coming from CMB+LSS, and BBN allow the existence of one sub-eV mass sterile
family but do not prefer extra fully thermalised sterile neutrinos in the eV-mass range since they  violate the
hot dark matter limit on the neutrino mass. However, the assumption of full thermalisation is not necessarily justified. In this paper, we
have studied the evolution of active and sterile neutrinos in the early universe in order to calculate the effective number of thermalized
species after $T \sim$ 1 MeV when active neutrinos have decoupled and slightly before BBN commences.
We have studied the amount of thermalisation for initial null and large ($L^{(a)} = 10^{-2}$) lepton asymmetry,
for a range of mass-mixing parameters and for both normal and inverted mass hierarchies.

Assuming null initial lepton asymmetry, we find that the assumption of full thermalisation is justified
for eV-mass sterile neutrinos with relatively large mixing (as suggested by short-baseline oscillation data).
This inevitably leads to tension between CMB+LSS data which prefers very light sterile neutrinos and Solar, reactor and short-baseline
data which prefers a mass around 1 eV or higher.

On the other hand, for large initial lepton asymmetries light sterile neutrinos are not (or only partially) thermalised for almost all the scanned
parameter space. This provides a loophole for eV sterile neutrinos to be compatible with CMB+LSS constraints. For lepton asymmetries around $10^{-2}$
almost no thermalisation occurs for the parameters preferred by Solar, reactor and short-baseline
data, and the sterile neutrinos would contribute very little to the current dark matter density.

One remaining open question neglected in this work is related to the impact of sterile neutrinos on BBN.
The $\nu_e$ and $\bar{\nu}_e$ flux distributions are affected by active-sterile conversions and they enter the weak rates regulating the neutron-proton equilibrium (see~\cite{Steigman:2007xt} for a review on the topic). Therefore the $^4$He abundance is sensitive to the presence of sterile families.
In particular $\delta N_{\rm eff} > 0$ and a less populated $\nu_e$ spectrum are both responsible for
increasing the freeze-out temperature of the ratio $n/p$ and therefore for a larger $^4$He abundance.

For small $L^{(a)}$  and the mixing parameters discussed here the active-sterile oscillations occur well before BBN commences, while
for large $L^{(a)}$ the active-sterile oscillations are no longer decoupled
from the active ones and can occur close to the BBN temperature. 
We refer the reader to \cite{Dolgov:2003sg} for a discussion of BBN constraints on the sterile sector, but also stress that for large values of $L^{(a)}$
any quantitative exclusion limits in mixing parameter space would require solving the full QKEs including all three active species. This is clearly beyond the scope of the present paper, but remains an interesting and important calculation.

Note added: After the initial version of this paper was finalised, a semi-analytic estimate of the BBN effect in the 3+1 scenario using the quantum rate equations has appeared~\cite{Mirizzi:2012we}.
%

\section*{Acknowledgments}
The authors are grateful to Georg G.~Raffelt for valuable discussions.
This work was partly supported by the Deutsche Forschungsgemeinschaft under the grant EXC-153
and by the  European Union FP7 ITN INVISIBLES (Marie Curie Actions, PITN-GA-2011-289442).
I.T.\ thanks the Alexander von Humboldt Foundation for support.

\appendix
\section{Location of the resonances}
\label{sec:resonances}

Imposing the resonance condition for neutrinos ($V_z = 0$) and for antineutrinos ($\overline{V}_z=0$),
one finds the locations of the resonances~\cite{Kainulainen:2001cb}. In order to make
explicit the $x$-dependence, we define
\begin{equation}
V_0 = \frac{\widetilde{V_0}}{x}\ \mathrm{and}\ V_1 = \widetilde{V_1}x\ .
\end{equation}
Introducing
\begin{equation}
\ell = \left\{
	\begin{array}{l l}
		\text{sign}[L^{(a)}]		& \quad\text{for particles} \\
		-\text{sign}[L^{(a)}] 	& \quad\text{for anti-particles}	
	\end{array}
\right.
\end{equation}
the resonance  conditions ($V_z = 0$ and $\overline{V}_z=0$) can be written
\begin{equation}
\widetilde{V_1}x^2 + \ell \fabs{V_L} x + \widetilde{V_0} = 0\ .
\end{equation}
We define $m\equiv \text{sign}[\delta m^2_s]$ and write the solution in the following way
\begin{align}
x_\text{res} &= x_0 \[ A\ell \pm \sqrt{A^2-m} \] \equiv x_0 F_{\ell m}^\pm\( A \) \label{eq:xres}\ ,
\end{align}
where we have defined
\begin{align}
x_0 &= \sqrt{\frac{m\widetilde{V_0}}{\widetilde{V_1}}}\ ,\\
A &= \frac{\fabs{V_L}}{2\sqrt{m\widetilde{V_0}\widetilde{V_1}}}\ ,\\
F_{\ell m}^\pm\(A\) &= \[ A\ell \pm \sqrt{A^2-m} \]\ .
\end{align}

Note that $x_0$ is always real and positive. In order to have a physical solution, $F_{\ell m}^\pm$  has to be real and positive. This
condition is satisfied for $F_{\pm 1,-1}^+(A)$ for any $A$ and $F_{+1,+1}^\pm(A)$ for $A \geq 1$. Thus, we always have two physical solutions when $m=-1$, one for particles and one for anti-particles. On the other hand, when $m=+1$ and $A\geq 1$, we have two resonances: when $\ell >0$, they occur for particles and when $\ell <0$, they occur for anti-particles being in both cases responsible for destroying the lepton number. These equations reproduce the ones reported in~\cite{Kainulainen:2001cb} when $m=-1$.

We can expand the solutions for small and large $L^{(a)}$:
\begin{subequations}
\label{eq:Fsymbol}
\begin{align}
F_{-1,-1}^+ &=-A+\sqrt{1+A^2} \simeq \left\{ \begin{array}{l l} 1-A+\frac{A^2}{2}-\cdots & A \rightarrow 0^+ \\
																															\frac{1}{2A} - \cdots & A\rightarrow \infty\end{array} \right.\\
F_{+1,-1}^+ &= \phantom{-}A+\sqrt{1+A^2} \simeq \left\{ \begin{array}{l l} 1+A+\frac{A^2}{2}-\cdots & A \rightarrow 0^+ \\
																															2A+\frac{1}{2A} - \cdots & A\rightarrow \infty\end{array} \right.\\
F_{+1,+1}^+ &=\phantom{-}A+\sqrt{A^2-1} \simeq \left\{ \begin{array}{l l} 1+\sqrt{2}\sqrt{A-1}+\cdots & A \rightarrow 1^+ \\
																															2A-\frac{1}{2A} - \cdots & A\rightarrow \infty\end{array} \right.\\
F_{+1,+1}^- &=\phantom{-}A-\sqrt{A^2-1} \simeq \left\{ \begin{array}{l l} 1-\sqrt{2}\sqrt{A-1}+\cdots & A \rightarrow 1^+ \\
																															\frac{1}{2A} - \cdots & A\rightarrow \infty\end{array}\right.
\end{align}
\end{subequations}
with $A$ and $x_0$ assuming the following expressions
\begin{align}
A &= \frac{6\zeta\para{3}}{\pi^3}   \sqrt{\frac{10}{7\sqrt{2}}}
\frac{ T \fabs{L^{(a)}} M_z \sqrt{G_F}}{\sqrt{\cos 2\theta_s \fabs{\delta m^2_s} \para{n_\nu + n_{\bar{\nu}}} g}} \label{eq:A}\\
& \simeq 7.28 \times 10^4\ T_\text{MeV} \frac{\fabs{L^{(a)}}}{\sqrt{\cos 2\theta_s \fabs{\delta m_s^2}_{\text{eV}^2} \para{n_\nu + n_{\bar{\nu}}} g}} \nonumber
\\
x_0 &= \frac{3}{\pi} \sqrt{\frac{5}{7\sqrt{2}}} \frac{M_z}{\sqrt{G_F}T^3} \sqrt{\frac{\cos 2\theta_s \fabs{\delta m^2_s}}{\para{n_\nu + n_{\bar{\nu}}} g}}
\simeq 1.81 \times 10^4\  T_\text{MeV}^{-3} \sqrt{\frac{\cos 2\theta_s \fabs{\delta m^2_s}_{\text{eV}^2}}{ \para{n_\nu + n_{\bar{\nu}}} g}}\ . \nonumber
\end{align}

For $L^{(a)}=10^{-2}$ we have $A\gg 1$ and therefore the lowest resonance will be
\begin{equation}
x_\text{res,low} \simeq \frac{x_0}{2A} = \frac{\pi^2 \cos 2\theta_s \fabs{\delta m^2_s}}{4\sqrt{2} \zeta\para{3} T^4 \fabs{L^{(a)}} G_F} \simeq 0.12 \frac{\cos 2\theta_s \fabs{\delta m^2_s}_{\text{eV}^2}}{T_\text{MeV}^4 \fabs{L^{(a)}}}\ . \label{xreslow}
\end{equation}
Note that $x_\text{res,low}$ is independent on the sign of the mass hierarchy and the total neutrino density.
Moreover, from the previous equation, we can extract the temperature at which the lowest resonance starts sweeping the bulk of the Dermi-Dirac distribution. This provides a good estimate of when resonant thermalisation sets in. For example, in the limit of large lepton number, assuming $x_\text{res,low} \simeq 0.1$, we find $T_\text{res,low} \simeq 3$~MeV for $(\delta m^2_s, \sin^2 2\theta_s)=(1\ \mathrm{eV}^2,10^{-2})$.
On the other hand, the higher resonance has no effect at all. In fact
\begin{equation}
x_\text{res,high} \simeq x_0\times 2A = \frac{180 \zeta\para{3} \fabs{L^{(a)}} M_z^2}{7\pi^2 T_\text{MeV}^2 \para{n_\nu + n_{\bar{\nu}}} g} \simeq 2.6 \times 10^{10} \frac{\fabs{L^{(a)}}}{\para{n_\nu + n_{\bar{\nu}}} g T_\text{MeV}^2}\ .
\end{equation}
Therefore, for $L^{(a)}=0.01$, $x_\text{res,high}$ will pass through the peak of the Fermi-Dirac distribution at $T \simeq 1$~GeV. At that temperature, the damping term is so strong that no oscillations occur and thermalisation is inhibited.

\section{Adiabatic approximation}
\label{sec:initialconditions}

The so-called ``adiabatic'' approximation was first introduced in~\cite{Foot:1996qc} and, under certain conditions, it allows one
to derive an approximate analytic solution of the QKE's. In this section, we closely follow the derivation from first principles of~\cite{Bell:1998ds}.
Such derivation assumes that the rate of repopulation ($\dot{P}_0$) vanishes, however the more careful analysis of~\cite{Lee:2000ej}, including a non-zero repopulation rate, turns out to give the same final formula for $P_y$. Therefore we choose to adopt the simpler derivation.

Assuming $\dot{P}_0 = 0$, Eq.~\eqref{qkep} can be written as a homogeneous matrix equation:
\begin{equation}
\frac{\di}{\di t}
\begin{bmatrix}
P_x \\ P_y \\P_z	
\end{bmatrix}
 =
\begin{bmatrix}
-D & -V_z & 0\\
V_z & -D & -V_x \\
0 & V_x & 0
\end{bmatrix}
P_z\ ,
\end{equation}
or using a vectorial notation
\begin{equation}
\frac{\di \vec{P}}{\di t} = \mathcal{K} \vec{P}\ .
\end{equation}

The matrix $\mathcal{K}$ can be diagonalised by a time-dependent matrix $\mathcal{U}$, such that $\mathcal{U}\mathcal{K}\mathcal{U}^{-1} = \mathcal{D}$. The matrix $\mathcal{U}$ defines an instantaneous diagonal basis through $\vec{Q} \equiv \mathcal{U} \vec{P}$ and,
in principle, the evolution equation for $\vec{Q}$ is non-trivial:
\begin{equation}
\label{eq:evoQ}
\frac{\di \vec{Q}}{\di t} = \mathcal{K} \vec{Q} - \mathcal{U} \frac{\di \mathcal{U}^{-1}}{\di t} \vec{Q}\ .
\end{equation}
However, if we assume that Eq.~\eqref{eq:evoQ} is dominated by the first term, the differential equation can be easily solved. This is the so-called
 ``adiabatic'' approximation and its applicability has been analysed thoroughly in~\cite{Bell:1998ds}. Quoting~\cite{Bell:1998ds}, it is applicable when
\begin{subequations}
\begin{align}
\frac{V_x}{\sqrt{D^2+V_z^2}} &\ll 1\ , \label{eq:smallVx}\\
T & \ll 3 \text{MeV}\ , \\
\fabs{\frac{\di L^{(a)}}{\di T_\text{MeV}}} & \ll 5 \times 10^{-11} T_\text{MeV}^4\ .
\end{align}
\end{subequations}

Equation~\eqref{eq:smallVx} is not easily stated as just a limit on temperature. If we are not close to the resonance and $V_z$ is dominated by $V_0$, we find $\tan2\theta_s \ll 1$ which is true for our parameter space. If we are close to the resonance, the criterion depends on $L^{(a)}$ through $x_\text{res}$
(see Appendix A for a discussion of the position of the resonances). Using Eqs.~(\ref{eq:xres},\ref{eq:Fsymbol}), we find
\begin{align}
L^{(a)}\gg 10^{-5} &: \frac{\fabs{V_x}}{D} \gtrsim \frac{\fabs{\delta m^2_s} \sin 2\theta_s }{C_a G_F^2 x_\text{res,low}^2 T^6} \sim 5\times 10^{11} \frac{T_\text{MeV}^2}{\fabs{\delta m_s^2}} \frac{\sin2\theta_s }{\cos^2 2\theta_s } L^{(a)2}\ ,\\
L^{(a)}\ll 10^{-5} &: \frac{\fabs{V_x}}{D} \gtrsim \frac{\fabs{\delta m_s^2} \sin 2\theta_s }{C_a G_F^2 x_0^2 T^6} \sim 50 \tan 2\theta_s\ .
\end{align}

For large $L^{(a)}$, we almost always break the approximation at the lowest resonance. But, since the resonance occurs
at a very low momentum, it would have no effect on the physics anyway. In principle, it could still affect numerics but we did not encounter problems on this particular front. For small $L^{(a)}$, we are safe for most of the parameter space and, as for large $L^{(a)}$, if the resonance is not sitting in a populated part of the Fermi-Dirac distribution, there should not be any impact on the physics from breaking this approximation slightly. This also applies to the third condition: If the resonance is not in the middle of a populated part of the distribution, we do not have a fast evolution of $L^{(a)}$ and the approximation is valid.

Equation~\eqref{eq:evoQ} can be formally solved by
\begin{equation}
Q_i(t) = \exp\( \int_{t_0}^t k_i(t') \di t'\) Q_i(t_0),
\end{equation}
where the $k_i$'s are the eigenvalues of $\mathcal{K}$. Expanding those to lowest order in $V_x$, we have
\begin{equation}
k_1 = -D+i V_z, \qquad k_2 = -D-i V_z, \qquad k_3 = -\frac{V_x^2D}{D^2+V_z^2}.
\end{equation}
Assuming that $D$ is large and $V_x$ satisfies Eq.~\eqref{eq:smallVx}, we find
\begin{equation}
Q_1(t) = Q_2(t) = 0, \quad Q_3(t) = Q_3(t_0).
\end{equation}
The adiabatic approximation allows us to relate $P_x$, $P_y$ and $P_z$ through
\begin{align}
\begin{bmatrix}
P_x(t)\\
P_y(t)\\
P_z(t)
\end{bmatrix}
&= \mathcal{U}^{-1}(t)
\begin{bmatrix}
Q_1(t)\\
Q_2(t)\\
Q_3(t)
\end{bmatrix}
=
\mathcal{U}^{-1}(t)
\begin{bmatrix}
0\\
0\\
Q_3(t_0)
\end{bmatrix}
=
Q_3(t_0) \vec{s_3(t)},
\end{align}
where $\vec{s_3}(t)$ is the third column in $\mathcal{U}^{-1}(t)$ which is also the normalised eigenvector corresponding to $k_3$. We have
\begin{equation}
\vec{s_3}(t) = N
\begin{bmatrix}
1\\
- (D+k_3)/V_z \\
-V_x (D+k_3)/(V_z k_3)
\end{bmatrix},
\end{equation}
with $N$ a normalisation constant. We can now relate $P_x$ and $P_y$ to $P_z$ to lowest order in $V_x$:
\begin{align}
P_x(t) &= \frac{V_x V_z}{D^2+V_z^2} P_z(t)\ , \\
P_y(t) &= -\frac{V_x D}{D^2+V_z^2} P_z(t)\ .
\end{align}
Substituting $V_z$ by $\overline{V}_z$ gives the corresponding relations for anti-particles.


\bibliographystyle{utcaps}

\bibliography{thermalisation_bib}

\end{document}